\documentstyle[twoside,12pt]{article}
\begin{document}
\begin{titlepage}
\begin{flushleft}
       \hfill                      USITP-94-04\\
       \hfill                       May 1994\\
       \hfill                      hep-th/9405195\\
\end{flushleft}
\vspace*{3mm}
\begin{center}
{\LARGE Strongly Topological Interactions \\
           of Tensionless Strings\\}
\vspace*{12mm}
{\large Bo Sundborg\footnote{E-mail: bo@vana.physto.se} \\
        {\em Institute of Theoretical Physics \\
        Box 6730\\
        S-113 85 Stockholm\\
        Sweden\/}\\}
\vspace*{25mm}
\end{center}

\begin{abstract}
The tensionless limit of classical string theory may be
formulated as a topological theory on the world-sheet. A vector density
carries geometrical information in place
of an internal metric. It is found
that path-integral quantization allows for
the definition of several, possibly
inequivalent quantum theories. String amplitudes
are constructed from  vector
densities with zeroes for each in- or out-going string. It is shown that
independence of a metric in quantum mechanical
amplitudes implies that the
dependence on such vector density zeroes is
purely topological. For  example,
there is no need for integration over their world-sheet positions.
\end{abstract}

\end{titlepage}

\section{Introduction}

String theory is characterized by amplitudes that can be represented as
sums or integrals over two-dimensional world-sheet geometries. The known
theories are
critical fundamental strings and two-dimensional Yang-Mills \cite{Gr1,GT}.
Whether Liouville or matrix-model strings are counted as a
separate class of theories or not, it has
proven very difficult to construct
string theories which are qualitatively different from the known
examples. There is an evident need for such theories if there is to be
progress in the long-standing problem of relating strings and strong
interactions.

Another problem which has proved surprisingly tough is the search
for a huge underlying symmetry of fundamental string theory. There are
many reasons to expect such symmetries: the low energy limits of string
theories are field theories with large
gauge symmetries, the massive string
modes appear in towers governed by symmetries of two-dimensional field
theories, high-energy behaviour and UV-divergences are mild even though
an infinite number of fields contribute to amplitudes, etc. Field theory
experience indicates that knowledge of
the full symmetry of a theory is useful. String theory should not be an
exception.

One approach to string symmetries has been to explore the expectation
that the symmetry is spontaneously broken in the present
formulation of the theory, but could become restored in some high-energy
limit. Symmetries between the exponentially small high-energy amplitudes
have been found by Gross \cite{Gr2} and
recently by Moore \cite{M}. A related but
different high-energy limit has
been studied at the non-interacting level
by taking the string tension to
zero [5-15]. In this limit {\em all\/} independent
relativistic momentum
invariants are assumed to be large compared to the tension, in contrast
to the fixed mass limits of Gross and Moore. To assess the value of the
limiting tensionless theory it is essential
to also consider interactions.

A direct way to see the importance of an interacting tensionless string
theory is to study the spectrum of the ordinary tensile string. It was
found in \cite{Su} that one-loop {\em mass level shifts are larger than
the level
separation\/} for high levels, no matter how small the coupling constant
is. In similar problems appearing in quantum mechanics, one turns to
quasi-degenerate perturbation theory. Taken over to string physics, this
recipe would mean that we start from the degenerate case with zero level
separation, then introduce interactions and finally a non-zero tension,
which causes the level splitting.
Zero separation is precisely the case of vanishing tension.

In the present article interactions are constructed in a geometric way
directly for the quantum tensionless string.
It should be stressed that this
procedure need not be equivalent to taking tensionless limits in
string amplitudes. Such an approach would first run into the problem of
properly identifying the limiting external states, since the spectra of
tensile and tensionless strings cannot be the same (because there is no
constant with dimensions in the tensionless case). By varying
external states while increasing the energy it is possible to
find high-energy limits for fundamental strings which are not
exponentially decreasing, in contrast to
standard fixed state limits \cite{Me}.
Moreover, there could be
qualitatively different high energy limits for different string theories.
For instance, in search of QCD strings one usually looks for a
different behaviour than in fundamental string amplitudes
\cite{Gree} in order to accommodate partonic structure. By defining an
interacting quantized tensionless
string theory per se, without recourse to limits of tensile amplitudes,
one has a chance to find alternative high energy theories. If one is
really lucky there could be background fields in these theories which
reintroduce tension and generate a new tensile
string theory.

In \cite{ILS,ILST} it was demonstrated by light-cone quantization that
preservation of the space-time conformal symmetry of tensionless strings
by quantization is incompatible with the naively expected spectrum.
Classically, the proof of conformal symmetry in the light-cone gauge
makes use of the reparametrization symmetry of the full theory, so the
anomaly is not to be taken lightly. Either conformal symmetry is broken
or the spectrum is quite different from what one gets by standard
quantization. The present geometrical approach does
not by itself favour any one
of these alternatives. When these techniques are developed further one
will however be able to deduce the spectrum, either by calculating the
partition function or by classifying the external states that couple
consistently to the surface geometry. The topological classification
found in this article is first step in this direction.

The article is organized as follows: in sect.\ 2 the action and the
geometrical formulation of interactions are described, in sect.\ 3 the
procedure for path integral quantization
is discussed and in sect.\ 4 the
topological nature of the quantum theory is examined. Section 5 contains
the methods for studying the space of vector density geometries
characterizing tensionless strings.
In sect.\ 6 these methods are applied
to the derivation of
the main results: absence of continuous moduli associated
to the insertion of external states
on the surface. Finally the conclusions
are summarized and set in perspective in sect.\ 7.

\section{The classical action and the geometry}

The geometric form of the classical action for the tensionless string
\cite{LST1} reads
\begin{equation}
        S_0
        ={\textstyle{1 \over 2}}\int {d^2}\xi \;V^aV^b\partial _aX\partial
        _bX .
        \label{tensionless action}
\end{equation}
To derive it one starts from the Nambu-Goto action, goes to phase
space and solves for the momenta to get a new configuration space action.
In the limit of vanishing tension it may be written as above.
Otherwise one obtains the ordinary tensile string action \cite{BDH,DZ}:
\begin{equation}
        S_T={\textstyle{T \over 2}}\int {d^2}\xi \,\sqrt {g\,}g^{ab}\partial
        _aX\cdot \partial _bX .
        \label{tensile action}
\end{equation}
One observes that $V^a V^b$ plays the role of $T\sqrt {g}g^{ab}$.
For the tensionless action to be diffeomorphism invariant $V^a$ should
thus transform as a vector density
and $X$ as a scalar. The infinitesimal
reparametrizations are
\begin{equation}
\begin{array}{rcl}
         \delta _\omega V^a &=& \omega ^b\partial _bV^a-V^b\partial
_b\omega ^a+{\textstyle{1 \over 2}}\partial _b\omega ^b V^a ,\\
        \delta _\omega X &=& \omega ^a\partial
_aX .
\end{array} \label{infinitesimal diffeomorphisms}
\end{equation}

The action is also invariant under spacetime conformal transformations.
While the Poincar\'{e} invariance is manifest, the conformal symmetry
acts  by rescaling $V^a$. For example, the conformal boosts are
\begin{eqnarray}
        X'^\mu & = & \left[ {X^\mu +b^\mu X^2} \right]
                    \left[ {1+2b\cdot X+b^2X^2} \right]^{-1} ,
        \label{conformal boosts on X} \\
        V'^a & = & V^a\sqrt {1+2b\cdot X+b^2X^2} .
        \label{conformal boosts on V}
\end{eqnarray}
In particular, note that conformal transformations taking a point $X_0$
to infinity also transforms a finite $V^a$ to zero.
One may thus consistently require that asymptotic string states
approaching infinity are accompanied by
vanishing vector densities on the
world-sheet.

{}From the tensionless action (eq. \ref{tensionless action}) one gets the
equations of motion
\begin{equation}
\begin{array}{rcl}
        0 & = & V^a\partial _aX\cdot \partial _bX ,\\
        0 & = & \partial _a\left( {V^aV^b\partial _bX} \right),
\end{array}
        \label{equations of motion}
\end{equation}
where the first equation gives the constraints analogous to the Virasoro
constraints of the tensile string, and the second equation may be
interpreted as the conservation of world-sheet momentum currents
{\em parallel
to $V^a$\/}.

A freely propagating string traces out
a cylindrical world sheet. On the
world sheet one can
draw flow lines whose tangents are
parallel to the vector density at each point.
Fig.\ 1 shows how diffeomorphism invariance
may be used to map the infinite
cylinder to a twice punctured sphere. The punctures correspond to
in-states or out-states, with the flow having a source or a sink at
the vector density zeroes coinciding with the punctures.
Vector densities on the
sphere with additional sources or sinks describe tree-level
{\em interacting\/} strings. Surfaces of higher genus represent emission
and
reabsorption of virtual strings, i.e.\ they give loop corrections.

In discussing the topologies of vector densities it is sometimes useful
to relate to concepts from the theory of dynamical systems or ordinary
differential equations \cite{AA}. To a {\em vector field\/} $ v^a $ one
may associate a differential equation
\begin{equation}
        {{d\xi ^a} \over {dt}}=v^a .
        \label{differential equation}
\end{equation}
One can associate a differential
equation to a vector density in the same
way at the price of not having a covariant
equation. A general coordinate
transformation then rescales the vector density.
Thus, the direction of a
solution trajectory behaves covariantly, but the time parameter $ t $ is
effectively rescaled in a position-dependent way. The most important
topological notions in dynamical systems theory are however independent
of the time parametrization, and therefore useful also for densities.

Zeroes of the vector density are then
equilibrium points. They are examples of limit
points, and there may also be
limit cycles. The topology of a vector density is reflected in the {\em
phase portrait\/} depicting flow-lines or
solution trajectories associated
with the density.

Fig.\ 2a
illustrates simple sources, saddles and sinks of a vector density on the
torus. It also exemplifies the Poincar\'{e}-Hopf theorem
\cite{PoincareHopf} relating the
numbers of different zeroes of vector fields on a genus $g$ surface. For
vector fields with only simple zeroes
\begin{equation}
        {\rm \#\, sources} - {\rm \#\, saddles} + {\rm \#\, sinks}
        = 2 (1 - g) .
        \label{index thm}
\end{equation}
Fig.\ 2b demonstrates that there are
smooth vector densities on the torus
without sources or sinks, but with stable and unstable cycles. We expect
that all possible flows should contribute to quantum mechanical
amplitudes. To determine their weights we have to define the path
integral over vector densities.

\section{The path integral}

The path integral measure may be defined by considering gaussian
integrals. The construction (see e.g.\ \cite{Sch})
requires that an inner
product on infinitesimal variations of fields is introduced,
and that the normalization
of the measure is specified \cite{Polch}.
Norm-preserving transformations
of
the fields become symmetries of the measure. For the tensile string this
approach was pioneered by Polyakov \cite{Poly}, Alvarez \cite{Al} and
Polchinski \cite{Polch}. It is reviewed in \cite{DP}, where further
references can be found.

The inner product should be reparametrization invariant. It should also
be local, which in this case means that
it has to be an integral over the
world-sheet of a function of the fields,
but not of their derivatives. The
inner product of two vector density variations then has to take the form
\begin{equation}
        \left\langle {{\delta V_1}}
        \mathrel{\left | {\vphantom {{\delta V_1}
{\delta V_2}}} \right. \kern-\nulldelimiterspace} {{\delta V_2}}
\right\rangle=\int {d^2\xi }\,g_{ab}(\xi )\delta V_1^a(\xi )\delta
V_2^b(\xi ) .
        \label{vector inner product}
\end{equation}

The function $g_{ab}$ has to transform as a tensor for the inner product
to be reparametrization invariant, and it has to be non-degenerate for
the inner product to be non-degenerate.
We may then regard it as a metric
on the world sheet, even though it does not appear in the action
(\ref{tensionless action}).

The metric allows for the construction of a new scalar
\begin{equation}
        \varphi \equiv g^{-1/2}g_{ab}V^aV^b .
        \label{scalar}
\end{equation}
The general reparametrization invariant and local inner product of
coordinate
variations is then
\begin{equation}
        {\left\langle {{\delta X_1}}
        \mathrel{\left | {\vphantom {{\delta X_1}
{\delta X_2}}} \right. \kern-\nulldelimiterspace} {{\delta X_2}}
\right\rangle=\int {d^2\xi }\,\sqrt g\kern 1pt\delta X_1\delta
X_2n(\varphi )} .
        \label{X inner product}
\end{equation}

The normalization of path integrals is fixed by
\begin{equation}
        \begin{array}{ccc}
           \int {DV^a}\,e^{-\int {d^2\xi \sqrt g}\mu _V(\varphi )}
e^{-\left\langle {{\delta V}} \mathrel{\left |
{\vphantom {{\delta V} {\delta V}}} \right. \kern-\nulldelimiterspace}
{{\delta V}} \right\rangle} = 1 , & &
                \int {DX}\,e^{-\int {d^2\xi \sqrt g}\mu _X(\varphi )}
 e^{-\left\langle {{\delta X}}
                \mathrel{\left | {\vphantom
{{\delta X} {\delta X}}} \right. \kern-\nulldelimiterspace} {{\delta X}}
\right\rangle} = 1
        \end{array}
        \label{normalization}
\end{equation}
where the functions \( \mu _V\) and \(\mu _X \)
encode our freedom in normalizing. They are
the most general expressions satisfying the principle of ultralocality
coined by Polchinski \cite{Polch}: the measure (and the ambiguity in it)
should be a reparametrization invariant
product of factors depending only
on the fields at individual world sheet points.

Three arbitrary functions \( \mu _V,\mu _X \) and \( n \) enter in the
definition of the measure. In the
case of non-vanishing tension the corresponding ambiguities are simply
constants, since no local scalar can be constructed solely from the
metric. If, however, we wish to make
contact with the classical theory we
should ask for path integrals independent of the metric. Then at least
part of the
ambiguity will be removed.

\section{Hyper-Weyl invariance}

Weyl invariance in the tensile string means that
the action (\ref{tensile
action}) is independent of the conformal
factor $\lambda$ when $g_{ab} \to
\lambda g_{ab}$. However, the path integral
measure in general depends on
the conformal factor through its definition in terms of an
inner product and through
regularization. Only for critical strings,
e.g.\ bosonic strings in flat
\( D=26 \) Minkowski space, does the
Weyl invariance survive quantization. In
this case one may choose to calculate in any convenient conformal factor
background.

The metric in the tensionless string is precisely analogous to the
conformal factor. It only enters upon
quantization, with the construction
of a measure. If metric independence,
characteristic of topological field
theories, is
maintained at quantization one may choose to calculate in any background
metric compatible with the surface topology. It is sometimes natural to
stress the analogy to Weyl-invariance by regarding metric independence
as an extension of this symmetry and
call it {\em hyper-Weyl invariance\/}.

In the Weyl invariant string the all-important geometry is the conformal
geometry which enters in the action,
and loop amplitudes can be expressed
as integrals over different conformal
geometries. Similarly the amplitudes
of the hyper-Weyl invariant tensionless
string are given by integrals (or
sums) over the geometries defined by the vector density \( V^a \).

It will be assumed in the following
that we are dealing with a hyper-Weyl
invariant quantization of the tensionless string. A simple one-loop
calculation in a flat background with a constant vector density suggests
that hyper-Weyl symmetry is non-anomalous, and the large freedom in the
definition of the path-integral
parametrized by the functions \( n, \mu_V \)
and \( \mu_X \) should help in avoiding anomalies in more complicated
backgrounds. I hope to settle this issue in future work, but for the
moment
the most pressing question is the nature of the vector density geometry
replacing the conformal geometry of
the tensile string. In particular, do
we have any right to call this theory topological? Even if it is
independent of the metric, the density \( V^a \) has appeared instead.

\section{Vector density dependence}

The action (\ref{tensionless action}) is diffeomorphism invariant.
We should fix a reparametrization gauge to select a
representative from each each equivalence class of
vector densities. Suppose that \( V^a \) has been fixed to
be a definite function on the surface. A general variation $\delta V^a$
may then be separated in two parts
\begin{equation}
        \delta V^a=(P_V\omega )^a+\delta m_j{{\partial V^a} \over {\partial
m_j}} ,
        \label{V variation}
\end{equation}
where the first term is just an infinitesimal diffeomorphism
(\ref{infinitesimal diffeomorphisms}), which may be written covariantly
\begin{equation}
        (P_V\omega )^a\equiv \delta _\omega V^a
        =\omega ^b\nabla _bV^a-V^b\nabla
_b\omega ^a+{\textstyle{1 \over 2}}\nabla _b\omega ^bV^a ,
        \label{covariant V reparametrization}
\end{equation}
now that we have introduced a metric. The second term accounts for
deformations of the geometry. There could be a space \(
\{V^a\}/ Diff \) of inequivalent vector
densities parametrized by moduli $m_i$.
In the presence of an inner product it is easy to identify
deformations caused by variations of the moduli. Since they cannot be
obtained
by a change of coordinates they may be chosen orthogonal to any
reparametrization.
Using the definition of adjoint operators this condition may be written
\begin{equation}
        0=\left\langle {{P_V\omega }}
        \mathrel{\left | {\vphantom {{P_V\omega }
{{{\partial V} \over {\partial m_j}}}}} \right.
\kern-\nulldelimiterspace}
{{{{\partial V} \over {\partial m_j}}}} \right\rangle\equiv \left\langle
{\omega } \mathrel{\left | {\vphantom {\omega  {P_V^{\dagger} {{\partial
V}
\over {\partial m_j}}}}} \right. \kern-\nulldelimiterspace}
{{P_V^{\dagger}
{{\partial V} \over {\partial m_j}}}} \right\rangle\,,\;\forall \omega .
        \label{orthogonality}
\end{equation}
Thus, possible deformations \( U^a \) of the
\( V^a \) geometry solve the
equation
\begin{equation}
        0 = (P_V^{\dagger} U)^a=\sqrt g\left[ {V^b\kern 1pt\nabla _bU^a-
{\textstyle{1 \over 2}}V_b\kern 1pt\nabla ^aU^b+
\nabla _bV^b\kern 1ptU^a+
{\textstyle{1 \over 2}}\nabla ^aV_b\kern 1ptU^b} \right]
        \label{zero-mode equation}
\end{equation}
and are zero-modes of
the operator \( P_V^{\dagger} \). In this equation it is important to
remember that \( U^a \) is a density and that accordingly there is an
extra term in the expression for the covariant derivative in terms of
Christoffel symbols. The dimension of the space of
geometries is the dimension of the
space of normalizable zero-modes, \({\rm Ker}
P_V^{\dagger} \).

Normalizability of \( U^a \) is needed
to make sense of path-integral
manipulations. However, the surfaces under
consideration
have punctures where external states
can be attached. For the ordinary string
this approach to amplitudes
is due to D'Hoker and Giddings \cite{DG}. Around
each puncture a small disc
is cut out. Then it is enough to have
normalizability in the remaining surface with arbitrarily small holes.
{\it This is the notion of normalizability used in the present paper.\/}
Though this definition of normalizability
does not give boundary conditions
for the behaviour of $U^a$ at zeroes (punctures), there are other
requirements. The {\it boundary conditions\/} for $U^a$
are obtained by demanding
that their
inner products with \( P_V \omega \) in
the orthogonality equation (\ref{orthogonality})
are finite
for all infinitesimal reparametrization symmetries $\omega ^a$ of the
{\it punctured\/} surface. The limits of these inner products should be
well defined as the sizes of the excluded discs approach zero.
A vector field $\omega ^a$ generates
a symmetry only if it
preserves the punctures, i.e.\ if it has
zeroes at all punctures. Thus we shall look
for normalizable solutions to the adjoint zero-mode equation
(\ref{zero-mode equation}), belonging to the space of vector densities
dual to the space of infinitesimal variations of the background vector
density $V^a$ due to such puncture-preserving diffeomorphisms.

Finding normalizable deformations \( U^a \)
is not a local issue. The norm
could diverge due to singularities anywhere on the world-sheet. We can,
however, limit the number of normalizable
\( P_V^{\dagger} \) zero-modes by
local arguments. A local feature of the vector density \( V^a \) at a
point $P$ could force zero-modes to have non-normalizable singularities
close to $P$. Then there are no deformations in a
neighbourhood of $P$, and there can be non-trivial normalizable
solutions on the surface only if their supports are separated from $P$
by a line where Taylor expansions of the zero-modes break down.

We have seen above in equation (\ref{index thm})
that local features like sinks,
sources and saddles of \( V^a \) are unavoidable
on world-sheets for interacting
tensionless strings. What are the consequences of such zeroes for the
existence of vector density deformations \( U^a \)?

\section{Topological interactions}
\label{Topological interactions}

We assume independence of the metric, which could
thus be taken to be flat in a neighbourhood of a zero.
We also assume that world-sheets have a
$ C^\infty $ differentiable structure so
that it makes sense to
study Taylor expansions of vector densities in any coordinate
chart. Generically the topology of the
flow close to the zero is determined
by the first terms of the expansion, but if they vanish the
topological classification of zeroes of vector fields in terms of
expansion coefficients is not complete \cite{AA}. However,
for non-zero real
analytic vector densities there are always some leading terms and in two
dimensions isolated analytic equilibrium points
can be classified topologically
(\cite{ALGM,AA}). Note that in contrast to the tensile case there is no
natural complex structure, and real analyticity of a function just means
that it is
Taylor-expandable with a non-vanishing radius of convergence.

Before discussing the classification we study reasonably
general but simple forms of the
densities \( V^a \). In polar coordinates:
\begin{equation}
        \begin{array}{ccc}
                V^r = r^{m+{\textstyle{1 \over 2}}}\cos(\delta + (p-1)\varphi)
,
  & &
                V^\varphi
                = r^{m-{\textstyle{1 \over 2}}}\sin(\delta + (p-1)\varphi) .
        \end{array}
                \label{zero forms}
\end{equation}
These zeroes behave simply under rotations; $p$ is a simple topological
invariant, the Poincar\'{e} index, which
measures how many times \( V^a \) winds
around the origin when its argument rotates once around the origin in a
small closed circuit, and $m$ is the order of the zero. If $m$ and $p$
are integers satisfying $m \geq 0$ and
$m \geq |p|$ these vector densities are real analytic, the factor
$\sqrt{r}$ being due to the density
factor in the transformation to polar
coordinates. The phase $\delta$ is irrelevant for $p\neq 1$. We then
absorb it in a shift of $\varphi$.

The zero-mode equation (\ref{zero-mode equation}) can be solved in such
backgrounds (\ref{zero forms}) by making the ansatz
\begin{equation}
        \begin{array}{ccc}
                U^r = r^k U_k^r(\varphi) ,  & &
                U^\varphi = r^{k-1} U_k^\varphi(\varphi) .
        \end{array}
        \label{deformation ansatz}
\end{equation}
The solution will be a sum of such terms, each obeying
\begin{eqnarray}
        0 & = & \sin ^2(\omega \varphi )\kern 1pt\;\ddot U_k^r+
             a\sin (2\omega \varphi )\kern 1pt\;\dot U_k^r+
          \left[ {b+c\sin ^2(\omega \varphi )} \right]\,U_k^r ,
        \label{ansatz equation} \\
        U_k^\varphi  &
        = & d\,\dot U_k^r+f\cot \kern 1pt(\omega \varphi )\;U_k^r
        \label{angular solution}
\end{eqnarray}
for $\delta = 0$.
Here $\omega = p - 1$, while $ a,b,c,d, e $ and $ f $ are somewhat
complicated rational
functions of $m,p$ and $k$. The final expressions given explicitly below
are much simpler.

For each value of $k$ there may be two,
one or zero normalizable solutions,
since the equation is second order.
To exclude solutions by local methods
we ask for which values of $m$ and $p$ all zero-modes are
non-normalizable. In those cases there can be no deformations of the
geometry. For $k > - \textstyle{3 \over 2} - m$
the radial solution satisfies
the boundary condition, but a singular angular dependence could cause a
divergence of the norm. Equation (\ref{ansatz equation})
has regular singular points
when $\sin{\omega \varphi} = 0$. The indicial equation
\begin{equation}
0=\zeta ^2+\left( {{a \over {2\omega }}-1} \right)\zeta +
{b \over {\omega ^2}}
        \label{indicial equation}
\end{equation}
then determines the exponents
\begin{eqnarray}
        \zeta _\pm & = & {1 \over {4\omega }}\left[ {(2\omega -a)\pm
        \sqrt {(2\omega -a)^2-16\,b}} \right]
        \nonumber \\
         & = &  -{1 \over {2\left( {p-1} \right)}}
         \left[{4m+3p+2k\mp (2m+p-1)}\right]
                    \label{exponents long}
\end{eqnarray}
in the power-law behaviour $U_k^\varphi \sim \varphi^\zeta $ close to
each singular point. The singular points actually represent singular
rays of constant $\varphi$ towards or away from the zero (\ref{zero
forms}). The condition for divergence of the norm
is $\zeta \leq - \textstyle{1
\over 2}$. If this inequality holds for both roots of the indicial
equation there can be no deformations.

Combining the conditions on angular
and radial behaviour of zero-modes we
find that there are no deformations of
regular vector densities with $p > 1$.
In contrast, there are always solutions
with locally convergent norms for
$p < 1$. The intermediate case $p = 1$ requires special attention. The
zero-mode equation (\ref{zero-mode equation})
should then be solved directly
without going via equation (\ref{angular solution}), which does not make
sense at $p = 1$.

For $p = 1$ the zero (\ref{zero forms}) with $\delta = 0$ describes the
kind of simple
source or sink that can be associated to asymptotic string states (Fig.\
1). Nonzero $\delta$ generically means that the flow spirals towards or
away from the zero. All such flows with $\cos{\delta} \neq 0$ are
topologically equivalent, and
solving (\ref{zero-mode equation}) one finds no normalizable solutions.
In contrast, the exceptional case $\cos{\delta} = 0$,
a `centre' where the flow
circles the zero without ever approaching or leaving, allows for
infinitely many solutions with locally convergent norms. To summarize,
there are {\em no continuous moduli\/}
for deformations of analytic vector
densities with
\begin{equation}
        \begin{array}{ccc}
                p \geq 1 , &  & \cos{\delta} \neq 0 .
        \end{array}
        \label{no moduli condition}
\end{equation}

These results tie in well with the
topological classification of isolated
zeroes of
analytic vector fields. Each topology except
the centre is represented by
vector densities consisting of a finite number of sectors of elliptic,
parabolic or hyperbolic type (see Fig.\ 3). If there are elliptic
sectors they always lie between two parabolic sectors. In this language,
we have found
that there could be deformations around centres or in hyperbolic
sectors. Zeroes with only parabolic and elliptic sectors cannot be
deformed and the dependence on such vector densities is truly
topological.

The zero-mode equations for the exceptional case of a centre (Fig.\ 3d)
appears to allow an infinite
number of deformations. Such infinitesimal deformations produce
spiralling vector densities of the same topology as the simple source or
sink. Centres should therefore be
regarded as unstable configurations, and
they do not give rise to any true continuous moduli.

For hyperbolic sectors the number of deformations can only be determined
by global considerations. The example on the torus in Fig.\ 2a
illustrates how moduli could be excluded: any non-zero deformation is a
zero-mode also close to the parabolic source (or the sink), where the
boundary conditions cannot be satisfied.

\section{Conclusions and outlook}

The geometry of bosonic tensionless
string theory is governed by a vector
density. In this article interactions have been introduced
by studying vector densities with zeroes on compact surfaces. We have
noted and parametrized a huge freedom
in path-integral quantization. At least part of this freedom is
presumably fixed when one imposes the condition that the quantum theory
be topological, in the sense of retaining the independence of a
two-dimensional metric which holds for the classical theory. If the
theory is not fixed completely by this requirement we obtain a class of
inequivalent high energy limits of string theory, some of which are not
necessarily limits of critical fundamental strings. The investigation of
the class of quantum topological
tensionless string theories is a crucial
one which I hope to return to.

On the assumption that the quantization
is independent of two-dimensional
metrics we have asked if the theory is what one may call `strongly
topological'. By this we mean that the dependence on the vector density,
the geometrical
object in the theory, should be given by its topology, and thus not
require the introduction of any continous moduli. This was
demonstrated for a large class of real analytic vector densities
(sect.\ \ref{Topological interactions}) with isolated limit points
surrounded only by so-called parabolic and
elliptic sectors. The case of ordinary sources and sinks (Fig.\ 1)
corresponding to classical in-coming and out-going strings is included.
There can be
moduli only if there are
regions on the surface to which certain zero-modes cannot be continued
analytically. The possible role of regions bounded by limit cycles for
this question is
under study. In this article it has been established that
there are no moduli associated with the external strings.

\bigskip

\begin{flushleft}
I wish to thank E.\ Aurell, H.\ Hansson, I.\ Kostov, U.\ Lindstr\"om,
M.\ Ro\v cek and G.\ Theodoridis for comments.
\end{flushleft}

\newpage
\begin{flushleft}
{\Large {\bf Figure captions}}
\end{flushleft}
        \begin{description}
                \item[1.]
                The diffeomorphism mapping an infinite cylinder to a
                punctured sphere.
                If there had been metrics on the surfaces, a Weyl
transformation
 would
                also have been needed to relate them.

                \item[2.]
                Opposite sides of the rectangles are identified to form a
torus.
                \begin{description}
                        \item[a)]  A source, a sink and two saddles on
                        the torus represent a
                        one-loop `self-energy correction'.

                        \item[b)]  One unstable and one stable limit cycle
            on the torus represent
                        a one-loop vacuum diagram.
                \end{description}

                \item[3.]
                Zeroes of different topology with Poincar\'{e} index $p$.
                \begin{description}
                        \item[a)] A zero with four hyperbolic sectors ($p=-1$).

                        \item[b)] A source with a single parabolic sector
($p=1$
).

                        \item[c)] A zero with four parabolic sectors
alternating

                        with four
                        elliptic sectors ($p=3$).

                        \item[d)] A centre ($p=1$).
                \end{description}
        \end{description}

\begin{thebibliography}{99}

\bibitem{Gr1}
D.J.\ Gross, {\it Nucl.Phys.\/} {\bf B400} (1993) 161.

\bibitem{GT}
D.J.\ Gross and W.\ Taylor, {\it Nucl.Phys.} {\bf B400} (1993) 181.

\bibitem{Gr2}
D.J.\ Gross, {\it Phys.Rev.Lett.} {\bf 60} (1988) 1229.

\bibitem{M}
G.\ Moore, {\it `Symmetries of the Bosonic String S-Matrix'\/},
Yale preprint
YCTP-P19-93 Oct.\ 1993, hep-th/9310026.

\bibitem{sc}
A.\ Schild, {\it Phys.Rev.} {\bf D16} (1977) 1722.

\bibitem{akul}
A.\ Karlhede and U.\ Lindstr\"om, {\it Class.Quant.Grav.}
{\bf 3} (1986) L73.

\bibitem{LST1}
U.\ Lindstr\"om, B.\ Sundborg and G.\ Theodoridis,
{\it Phys.Lett.} {\bf B253} (1991) 319;
{\it Phys.Lett.} {\bf B258} (1991) 331.

\bibitem{roli}
M.\ Ro\v cek and U.\ Lindstr\"om,{\it Phys.Lett.} {\bf B271} (1991) 79.

\bibitem{zh}
A.A.\ Zheltukhin,{\it Sov.J.Nucl.Phys.} {\bf 48} (1988) 375.

\bibitem{banerual}
A.\ Barcelos-Neto and M.\ Ruiz-Altaba,{\it Phys.Lett.}
{\bf B228} (1989) 193.

\bibitem{liraspsr}
F.\ Lizzi, B.\ Rai, G.\ Sparano and A.\ Srivastava,
{\it Phys.Lett.} {\bf B182} (1986) 326.

\bibitem{gararual}
J.\ Gamboa, C.\ Ramirez and M.\ Ruiz-Altaba,
{\it Nucl.Phys.} {\bf B338} (1990) 143.

\bibitem{ILS}
J.\ Isberg, U.\ Lindstr\"om and B.\ Sundborg,
{\it Phys.Lett.\/} {\bf B293} (1992) 321.

\bibitem{ILST}
J.\ Isberg, U.\ Lindstr\"om, B.\ Sundborg and G.\ Theodoridis, {\it
Nucl.Phys.\/} {\bf B411} (1994) 122.

\bibitem{BSTV}
I.A.\ Bandos, D.P.\ Sorokin, M.\ Tonin and
D.V.\ Volkov, {\it Phys.Lett.\/} {\bf B319}
(1993) 445.

        \bibitem{Su}
        B.\ Sundborg, {\it Nucl.Phys.\/} {\bf B319} (1989) 415.

        \bibitem{Me}
        P.F.\ Mende, {\it `High-Energy String Collisions
in a Compact Space'\/}, Brown
        preprint BROWN-HET-929, Jan.\ 1994, hep-th/9401126.

        \bibitem{Gree}
        M.B.\ Green, {\it Phys.Lett.\/} {\bf B266} (1991) 325.

        \bibitem{BDH}
        L.\ Brink, P.\ Di Vecchia and P.S.\ Howe,
{\it Phys.Lett.\/} {\bf B65} (1976)
        435.

        \bibitem{DZ}
        S.\ Deser and B.\ Zumino, {\it Phys.Lett.\/} {\bf B65} (1976) 369.

    \bibitem{AA}
        V.I.\ Arnold and Yu.S.\ Il'yashenko,
{\it Ordinary differential equations\/},
        in: Dynamical systems I, Vol.\ I of Encyclopaedia of Mathematical
        Sciences, eds. D.V.\ Anosov and V.I.\ Arnold (Springer, 1988) p.\ 1.

        \bibitem{ALGM}
        A.A.\ Andronov, E.A.\ Leontovich, I.I.\ Gordon and A.G.\ Maier,
        {\it Qualitative theory of second order dynamic systems\/}
        (Wiley, Israel
        program for scientific translations, 1973).

        \bibitem{PoincareHopf}
        Y.\ Choquet-Bruhat and C.\ DeWitt-Morette,
        {\it Analysis, manifolds and physics, Part II: 92 Applications\/}
        (North-Holland, 1989).

        \bibitem{Sch}
        A.S.\ Schwarz, {\it Commun.Math.Phys.\/} {\bf 67} (1979) 1.

        \bibitem{Polch}
        J.\ Polchinski, {\it Commun.Math.Phys.\/} {\bf 104} (1986) 37.

        \bibitem{Poly}
        A.M.\ Polyakov, {\it Phys.Lett.\/} {\bf B103} (1981) 207.

        \bibitem{Al}
        O.\ Alvarez, {\it Nucl.Phys.\/} {\bf B216} (1983) 125.

        \bibitem{DP}
        E.\ D'Hoker and D.H.\ Phong, {\it Rev.Mod.Phys.\/} {\bf 60}
        (1988) 917.

    \bibitem{DG}
        E.\ D'Hoker and S.B.\ Giddings, {\it Nucl.Phys.\/} {\bf B291} (1987)
90.

\end{thebibliography}
\end{document}